\documentclass{pasj00}
\usepackage[dvips]{graphicx}
\SetRunningHead{Astronomical Society of Japan}{Magnetically Supported
Accretion Disk} 
\title{Steady Models of Optically Thin, Magnetically Supported  
Black Hole Accretion Disks}
\author{Hiroshi \textsc{Oda}}
\affil{Graduate School of Science and Technology, Chiba University,
1--33 Yayoi-Cho, Inage-ku, Chiba 263--8522}
\email{oda@astro.s.chiba-u.ac.jp}
\author{Mami \textsc{Machida}}
\affil{Division of Theoretical Astronomy, National Astronomical
Observatory of Japan, 2--21--1 Osawa,\\ 
Mitaka, Tokyo 181--8588}
\email{mami@th.nao.ac.jp}
\author{Kenji E. \textsc{Nakamura}}
\affil{Department of Sciences, Matsue National College of Technology,
14--4 Nishiikuma-Cho,\\ 
Matsue, Shimane 690--8518}
\email{nakamrkn@matsue-ct.jp}
\and
\author{Ryoji \textsc{Matsumoto}}
\affil{Department of Physics, Faculty of Science, Chiba University,
1--33 Yayoi-Cho, Inage-ku, Chiba 263--8522}
\email{matumoto@astro.s.chiba-u.ac.jp}

\Received{$\langle$reception date$\rangle$}
\Accepted{$\langle$acception date$\rangle$}
\Published{$\langle$publication date$\rangle$}
\begin{document}
\maketitle

\begin{abstract}
We obtained steady solutions of optically thin, single temperature,
 magnetized black hole accretion disks assuming thermal bremsstrahlung
 cooling. Based on the results of
 3D MHD simulations of accretion disks, we assumed that the magnetic
 fields inside the disk are turbulent and dominated by azimuthal
 component. We decomposed magnetic fields into an azimuthally averaged
 mean field and fluctuating fields. We also assumed that the azimuthally averaged
 Maxwell stress is  proportional to the total pressure. 
 The radial advection rate of the azimuthal magnetic
 flux $\dot \Phi$ is prescribed as being proportional to $\varpi^{- \zeta}$,
 where $\varpi$ is the radial coordinate and $\zeta$ is a parameter
 which parameterizes the radial variation of $\dot \Phi$. We found that
 when accretion rate $\dot M$ exceeds the threshold for the onset of the
 thermal instability, a magnetic pressure
 dominated new branch appears.
Thus the thermal equilibrium curve of optically thin disk has a
 'Z'-shape in the plane of surface
 density and temperature. This
 indicates that as the mass accretion rate increases, a gas pressure
 dominated optically thin hot accretion disk undergoes a transition to a
 magnetic pressure dominated, optically thin cool
 disk. This disk corresponds to the X-ray hard, luminous disk in
 black hole candidates observed
 during the transition from a low/hard state to a high/soft state. We also obtained global
 steady transonic solutions containing such a transition layer.      
\end{abstract}

\section{Introduction}

Optically thin, hot accretion disks have been studied to explain the
X-ray hard state (low/hard state) of black hole
candidates. \citet{thor75} proposed that hard X-rays from Cyg X-1 are
produced in the
inner optically thin hot disks. \citet{shib75} studied the
structure and stability of optically thin hot accretion
disks. \citet{eard75} and \citet{shap76} constructed a model of
optically thin, two temperature accretion disks
in which ion temperature is higher than the electron
temperature. \citet{ichi77} pointed out the importance of advection of
energy in hot accretion flows and obtained steady models of
optically thin disks by equating the radial heat advection $Q_{\rm adv}$ and
the viscous heating $Q_{\rm vis}$. Such flows, which now are called
advection dominated accretion flows (ADAFs) or radiatively inefficient
accretion flows (RIAFs), were studied
extensively by Narayan, Yi (1994,1995) and \citet{abrm95}. 
\citet{abrm95} obtained local thermal equilibrium curves of 
optically thin accretion disks by solving $Q_{\rm vis}=Q_{\rm
adv}+Q_{\rm rad}$, where $Q_{\rm rad}$ is the radiative cooling
rate. They showed that when the 
accretion rate exceeds a critical accretion rate, optically thin
solution disappears. Above this critical accretion rate, the hard X-ray
emitting optically thin hot accretion disk undergoes a transition to an
optically thick cool disk which emits soft X-rays. The latter corresponds
to the high/soft state (or thermal state) of black hole
candidates.

The importance of the magnetic fields on transition between 
an optically thin hot disk and an optically thick cool disk was discussed
by \citet{ichi77}. He suggested that magnetic pressure $p_{\rm mag}$ is limited
below the gas pressure $p_{\rm gas}$ because magnetic flux will escape from the
disk by buoyancy. However, the buoyant escape of the magnetic flux can
be suppressed in strongly magnetized disks.  \citet{shib90} carried out
magnetohydrodynamic (MHD) simulations of the buoyant escape of the magnetic
flux due to the Parker 
instability (\cite{park66}) and showed that when the disk is dominated by
magnetic pressure, it can stay in low-$\beta$ state ($\beta = p_{\rm
gas}/p_{\rm mag} < 1$) because the growth rate of the Parker instability
is reduced in low-$\beta$ disks due to the magnetic
tension. \citet{mine95} suggested that low-$\beta$ disk emits hard
X-rays. \citet{pari03} obtained steady solutions
of optically thick, magnetic pressure dominated disk but they did not
consider the optically thin solution including advective
cooling. \citet{bege06} constructed a model of optically thick
low-$\beta$ disk including radiation pressure. 

Since magnetic fields contribute both to the angular momentum transport
and disk heating, we need to understand how magnetic fields are
amplified and maintained in the disk. In conventional theory of
accretion disks (e.g., \cite{shak73}), phenomenological
$\alpha$-viscosity is invoked. However, the physical mechanism which
enables angular momentum transport efficient enough to explain the
activities of accretion powered sources such as dwarf nova was unknown
until \citet{balb91} pointed out the importance of the magneto-rotational
instability (MRI) in accretion disks. 
MRI can excite and maintain magnetic
turbulence. The Maxwell stress generated by the turbulent magnetic fields
efficiently transports angular momentum and enables accretion of the
disk material. 
The growth and the saturation of MRI has been studied by local
three-dimensional MHD simulations (e.g., \cite{hawl95},
\yearcite{hawl96}; \cite{mats95};
\cite{bran95}; \cite{sano01}; \cite{sano04}) and by global
three-dimensional MHD simulations (e.g.,
\cite{mats99}; \cite{hawl00}; \cite{mach00}; \cite{hawl01};
\cite{hawl02}; \cite{mach03}). These global
three-dimensional MHD simulations of radiatively inefficient accretion
disks indicate that the amplification of magnetic fields saturates when
$\beta \sim 10$ except in the plunging region of black hole accretion
disk and in the disk corona, and that the disk approaches to a
quasi-steady state. In this quasi-steady state, the average ratio of
the Maxwell stress to gas pressure $\alpha_{\rm SS} $, which corresponds to
the $\alpha$-parameter in the conventional accretion disk theory
(\cite{shak73}), is $\sim 0.01 - 0.1$ (e.g., \cite{hawl00};
\cite{hawl01}; \cite{mach03}).

RXTE observations of state transitions in galactic black hole candidates
revealed that some black hole candidates (e.g., GX339-4, XTE J1550-564, 
and XTE J1859+226) stay in X-ray
hard states even when their luminosities exceed 20\% of the
Eddington luminosity (e.g., \cite{done03}; \cite{gier06}). When $\alpha
\ltsim 0.1$ as indicated by global three-dimensional MHD
simulations of black hole 
accretion flows and the energy conversion
efficiency $\eta_{\rm e} \sim 0.1$, their luminosities are higher than the
critical luminosity above which optically thin steady hot
solutions disappear.

Recently, \citet{mach06} carried out global
three-dimensional MHD simulations of black hole accretion disks 
including optically thin radiative cooling. They started the simulation
from a radiatively inefficient torus threaded by weak toroidal magnetic fields.
As MRI grows, an optically thin, hot accretion disk is formed by
transporting angular momentum efficiently. They found by simulation
including thermal bremsstrahlung cooling that when the
density of the accretion disk exceeds the critical density, 
cooling instability takes place in the disk. The disk shrinks
 in the vertical direction with almost conserving the toroidal magnetic
flux. They demonstrated that as the disk shrinks, magnetic pressure 
exceeds the gas pressure because the magnetic pressure increases due to flux
conservation, meanwhile the gas pressure decreases due to cooling.  
When the magnetic pressure becomes dominant, the disk stops
shrinking in the vertical direction because the magnetic pressure supports
the disk. They found that the disk evolves toward a quasi-equilibrium
cool state. During this transition from hot state to cool state, the
disk remains optically thin. They explained this quasi-steady state by
equating the magnetic heating rate $ Q^{+}$ and the radiative cooing
rate $Q_{\rm rad}$ assuming that the Maxwell stress is proportional to the
total pressure and that the azimuthal magnetic flux $\langle B_{\varphi}
\rangle H$
is conserved during the transition, where $ \langle B_{\varphi} \rangle
$ is the mean azimuthal magnetic field, and $H$ is the half thickness of the disk.

The purpose of this paper is to answer why hard
X-ray emitting optically thin disks exist above the critical
luminosity.
We show steady solutions of optically thin black
hole accretion disks by taking into account the magnetic fields, radiative
cooling, and advection of energy and
magnetic fields. 
In section 2, we present basic equations. In section 3,
we obtain local thermal equilibrium curves. We present the global
steady solutions in section 4. Section 5 is devoted for summary
and discussion.

\section{Models and Assumptions}
\subsection{Basic Equations}
We extended the basic equations for one-dimensional steady, optically
thin black hole accretion flows (e.g., \cite{kato98}) by incorporating
the magnetic fields. We adopt cylindrical coordinates
$(\varpi,\varphi,z)$. General relativistic effects are
simulated by using the pseudo-Newtonian potential $\psi =
-GM/(r-r_{\rm s})$ (\cite{pacz80}), where $G$ is the
gravitational constant, $M$ is the mass of the black hole (we assume $M
= 10 \MO$ in this paper), $r = (\varpi^2 + z^2)^{1/2}$, and $r_{\rm s} =
2GM/c^2 $ is the Schwarzschild radius.

We start from the resistive MHD equations

\begin{eqnarray}
 \frac{\partial \rho}{\partial t} + \nabla \cdot \left( \rho
  \mbox{\boldmath{$v$}} \right) = 0 ~,
 \label{eq:vec_continuity}
\end{eqnarray}

\begin{eqnarray}
 \rho \left[ \frac{\partial \mbox{\boldmath{$v$}}}{\partial t} + \left(
 \mbox{\boldmath{$v$}}   
 \cdot \nabla \right) \mbox{\boldmath{$v$}}
      \right] = - \rho \nabla \psi - \nabla p_{\rm gas} +
 \frac{\mbox{\boldmath{$j$}} \times \mbox{\boldmath{$B$}}}{c} ~,
 \label{eq:vec_mom}
\end{eqnarray}

\begin{eqnarray}
 \frac{\partial \left( \rho \epsilon \right)}{\partial t} + \nabla \cdot
  \left[ \left(
 \rho \epsilon + p_{\rm gas}\right) \mbox{\boldmath{$v$}} \right] - \left(
 \mbox{\boldmath{$v$}} \cdot \nabla \right) p_{\rm gas} = q^{+} - q^{-} ~,
 \label{eq:vec_energy}
\end{eqnarray}

\begin{eqnarray}
 \frac{\partial \mbox{\boldmath{$B$}}}{\partial t} = \nabla \times \left( \mbox{\boldmath{$v$}}
 \times \mbox{\boldmath{$B$}} - \frac{4 \pi }{c}{\eta} \mbox{\boldmath{$j$}} \right) ~,
 \label{eq:vec_ind}
\end{eqnarray}
where $\rho$ is the density, $\mbox{\boldmath{$v$}}$ is the velocity, $\mbox{\boldmath{$B$}}$
is the magnetic field, $\mbox{\boldmath{$j$}} = c \nabla \times \mbox{\boldmath{$B$}} / 4 \pi$
is the current density, $p_{\rm gas} = \Re \rho T / \mu$ is the gas
pressure, $T$ is the temperature,  
$\epsilon = 3 \Re T / 2 \mu $ is the internal gas energy, $\Re$ is
the gas constant, $\mu$ is the mean molecular weight (we assumed to be
$0.617$), $q^{+}$ is the heating rate, $q^{-}$ is the radiative cooling
rate, and $\eta$ is the resistivity. 

Global three-dimensional MHD simulations (e.g., \cite{mats99},
\cite{hawl00}, \cite{mach00}; \cite{hawl01}, \yearcite{hawl02};
\cite{mach03}; \cite{mach04}, \yearcite{mach06} ) showed that in radiatively inefficient accretion disks
the amplified turbulent magnetic fields saturates when $\beta =
p_{\rm gas}/p_{\rm mag} \sim 10$ except in the plunging region and in the disk
corona.  Inside the disk, the azimuthal component of
magnetic field dominates the poloidal component.

\citet{mach06} carried out three-dimensional MHD simulation of
black hole accretion disks including optically thin radiative cooling
and showed that when the mass accretion rate exceeds the threshold for
the onset of the cooling instability, a magnetically supported disk is
formed. The magnetic pressure exceeds the
gas pressure because magnetic fields are amplified due to the vertical
contraction of the disk, meanwhile the gas pressure decreases due
to cooling.

We have to take into account magnetic fields to study
the evolution of accretion disks in this regime.
We decompose the magnetic fields into the
axisymmetric toroidal component (mean field) $\mbox{\boldmath{$\bar{B}$}} =
\langle B_{\varphi} \rangle \mathbf{{\hat{e}_{\varphi}}}$ and the
fluctuating fields $\delta \mbox{\boldmath{$B$}} = \delta B_{\varpi}
\mathbf{\hat{e}_{\varpi}} + \delta B_{\varphi} \mathbf{\hat{e}_{\varphi}} +
\delta B_{z} \mathbf{\hat{e}_{z}}$ and decompose the
velocity into the mean velocity $\mbox{\boldmath{$\bar{v}$}} =
(v_{\varpi}, v_{\varphi}, v_{z})$ and
fluctuating velocity $\delta \mbox{\boldmath{$v$}}$. Here $\langle ~~ \rangle$
denotes the azimuthal average and $B_{\varpi} = \delta B_{\varpi}$,
$B_{\varphi} = \langle B_{\varphi} \rangle + \delta B_{\varphi}$, and $B_{z} =
\delta B_{z}$. We assume that the
fluctuating components vanish when azimuthally averaged ($\langle \delta
\mbox{\boldmath{$v$}} \rangle = \langle \delta \mbox{\boldmath{$B$}}
\rangle = 0$). Thus, the
azimuthally averaged magnetic field and velocity are $\langle
\mbox{\boldmath{$B$}}
\rangle =  \langle B_{\varphi} \rangle \mathbf{\hat{e}_{\varphi}}$ and
$\langle \mbox{\boldmath{$v$}} \rangle = \mbox{\boldmath{$\bar{v}$}}$, respectively.

Let us derive azimuthally averaged equations assuming that the disk is
in a steady
state. We assume that $|\langle B_{\varphi} \rangle + \delta
B_{\varphi}| \gg |\delta B_{\varpi}|$, $|\delta B_{z}|$. By azimuthally
averaging equations (\ref{eq:vec_continuity}) - (\ref{eq:vec_energy})
and ignoring the second order term of $\delta B_{\varpi}$ and $\delta
B_{z}$,
we obtain

\begin{equation}
 \label{eq:continuity}
 \frac{\partial}{\varpi \partial \varpi} \left( \varpi \rho
 v_{\varpi} \right) +
 \frac{\partial}{\partial z} \left( \rho v_{z} \right) = 0 ~,
\end{equation}

\begin{equation}
 \label{eq:mom_pi}
 \rho v_{\varpi} \frac{\partial v_{\varpi}}{\partial \varpi} + \rho
  v_{z} \frac{\partial v_{\varpi}}{\partial z} - \frac{\rho
  v_{\varphi}^{2}}{\varpi} = - \rho \frac{\partial \psi}{\partial
  \varpi} - \frac{\partial p_{\rm tot}}{\partial \varpi} +
  \frac{\langle B_{\varphi}^2 \rangle}{4 \pi \varpi } ~, 
\end{equation}

\begin{eqnarray}
  \label{eq:mom_phi}
 \rho & v_{\varpi} & \frac{\partial v_{\varphi}}{\partial \varpi} + \rho
  v_{z} \frac{\partial v_{\varphi}}{\partial z} + \frac{\rho
  v_{\varpi} v_{\varphi}} {\varpi} \nonumber \\ 
&=&  \frac{1}{{\varpi}^{2}}
  \frac{\partial}{\partial \varpi} \left[ {\varpi}^{2} \frac{ \langle
				    B_{\varpi} 
B_{\varphi} \rangle}{4\pi} \right] + \frac{\partial}{\partial z}
  \left( \frac{\langle B_{\varphi} B_{z} \rangle}{4 \pi}\right)~,
\end{eqnarray}

\begin{equation}
  \label{eq:mom_z}
 \rho v_{\varpi} \frac{\partial v_{z}}{\partial \varpi} + \rho
  v_{z} \frac{\partial v_{z}}{\partial z} = - \frac{\partial \psi}{\partial z}
  - \frac{1}{\rho} \frac{\partial p_{\rm tot}}{\partial z} ~,
\end{equation}

\begin{eqnarray}
  \label{eq:energy}
 \frac{\partial}{\partial \varpi} \left[ \left( \rho \epsilon +
					       p_{\rm gas} \right)
 v_{\varpi}\right]
 & + & \frac{v_{\varpi}}{\varpi} \left( \rho \epsilon +
 p_{\rm gas} \right) + \frac{\partial}{\partial z} \left[
 \left( \rho \epsilon + p_{\rm gas} \right) v_{z}\right] \nonumber \\
& - & v_{\varpi} \frac{\partial p_{\rm gas}}{\partial \varpi}
- v_{z} \frac{\partial p_{\rm gas}}{\partial z} = q^{+} - q^{-} ~,
\end{eqnarray}
where $p_{\rm tot} = p_{\rm gas} + p_{\rm mag}$ is the total
pressure and $ p_{\rm mag} = \langle B_{\varphi}^{2} \rangle /8 \pi$ is the
azimuthally averaged magnetic pressure. 

Global three-dimensional MHD simulations showed that, in radiatively
inefficient disks, azimuthally averaged total stress
inside the disk is dominated by the $\varpi \varphi$-component of the
Maxwell stress $\langle t_{\varpi \varphi} \rangle
\sim \langle B_{\varpi} B_{\varphi} /4 \pi
\rangle$ and that the average ratio of the Maxwell stress to the magnetic
pressure, $\alpha_{\rm m} \equiv - \langle B_{\varpi} B_{\varphi} /4 \pi
\rangle / \langle p_{\rm mag} \rangle $, is nearly constant ($\sim 0.2 -
0.5$ ; e.g., \cite{hawl01}, \yearcite{hawl02}; \cite{pess06}). Since $\beta \sim 10$ inside the disk except in the plunging
region, $\alpha_{\rm B} \equiv - \langle B_{\varpi} B_{\varphi} / 4 \pi
\rangle / p_{\rm tot} \sim 0.01 - 0.1$ (e.g.,
\cite{hawl01}). \citet{mach06} showed that $\alpha_{\rm B} \sim 0.05 -
0.1$ in cooling dominated magnetic pressure supported disk.
According to their numerical results, we assume that the azimuthally
averaged $\varpi \varphi$ component of the Maxwell stress is
proportional to the total pressure ($ \langle B_{\varpi} B_{\varphi} / 4
\pi \rangle =
- \alpha_{\rm B} p_{\rm tot}$). Thus, 
equation (\ref{eq:mom_phi}) becomes

\begin{eqnarray}
 \rho & v_{\varpi} & \frac{\partial v_{\varphi}}{\partial \varpi} + \rho
  v_{z} \frac{\partial v_{\varphi}}{\partial z} + \frac{\rho
  v_{\varpi} v_{\varphi}} {\varpi} \nonumber \\
&=& \frac{1}{{\varpi}^{2}}
  \frac{\partial}{\partial \varpi} \left[ {\varpi}^{2} \left(-
 \alpha_{\rm B} p_{\rm tot} \right) \right] + \frac{\partial}{\partial z}
  \left( \frac{\langle B_{\varphi} B_{z} \rangle}{4 \pi}\right)~.
\end{eqnarray}

\begin{figure}[h]
 \begin{center}
  \includegraphics[width=80mm]{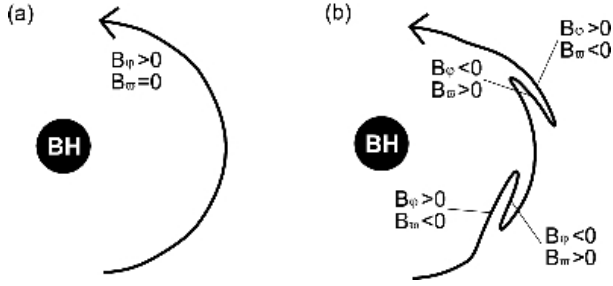}
  \caption{A schematic picture of a magnetic field line inside the
  accretion disk. (a) Axisymmetric disk threaded by mean azimuthal
  magnetic fields. (b) Turbulent disk with mean azimuthal magnetic
  fields and fluctuating magnetic fields. The azimuthal average of the radial component of the
  magnetic field, $\langle B_{\varpi} \rangle$, is small because positive and
  negative $B_{\varpi}$ cancel out. On the other hand, the
  azimuthal average of the product
  of the radial and toroidal component, $\langle B_{\varpi} B_{\varphi}
  \rangle$, has a large negative value because $B_{\varpi} B_{\varphi}$
  does not change sign when magnetic fields are deformed by nonlinear
  growth of MRI.
}
  \label{line}
 \end{center}
\end{figure}

We note that in the
turbulent accretion disk, the average of the radial component of
magnetic fields, $\langle B_{\varpi} \rangle$, is small because positive
and negative $B_{\varpi}$ cancel out. On the other hand, the
product of the radial and the toroidal components, $B_{\varpi} B_{\varphi}$,
does not change sign in the non-linear state (see figure
\ref{line}). Thus, we can not neglect  $\langle B_{\varpi} B_{\varphi}
\rangle$ term.

\subsection{Heating and Cooling Rates}
The other key factor is the heating rate expressed as
$q_{\rm vis}^{+}= t_{\varpi \varphi} \varpi \left(d \Omega /d \varpi
\right)$ in conventional theory, where $\Omega$ is the angular
velocity. Meanwhile, three-dimensional MHD simulations indicate that
the total dissipative heating rate (a major part of dissipation
is due to the thermalization of magnetic energy via magnetic reconnection)
is $q^{+} \sim \langle B_{\varpi} B_{\varphi}/4 \pi
\rangle \varpi \left( d \Omega / d \varpi\right) $ and that this dissipation
occurs throughout the disk (e.g., \cite{hiro06}).

We assume the magnetic heating as the
heating mechanism in the disk and set the heating term as follows:

\begin{equation}
 q^{+} = \frac{\langle B_{\varpi} B_{\varphi} \rangle}{4 \pi} \varpi
  \frac{d\Omega}{d\varpi} = - \alpha_{\rm B} p_{\rm tot} \varpi
  \frac{d\Omega}{d\varpi} ~.
\end{equation}

As the cooling mechanism, we assume the radiative cooling
by the optically thin thermal bremsstrahlung emission. Hence, the
cooling rate is expressed as,
\begin{equation}
q^{-} = 6.2 \times 10^{20} {\rho}^{2} {T}^{1/2} ~{\rm ergs~s^{-1}~cm^{-3}} .
\end{equation}

\subsection{Vertically Integrated Equations}
We assume that the temperature $T$, 
$\beta \left(= p_{\rm gas}/p_{\rm mag} \right)$,
radial velocity $v_\varpi$, and specific angular
momentum $\ell (= \varpi v_{\varphi})$ are independent of
$z$. \citet{mach00} and \citet{mill00} showed that the magnetic loops
emerge from the disk
and form coronal magnetic loops. In this paper, we
focus on the disk and ignored the low density and low $\beta$ corona. We assume
hydrostatic balance in the vertical direction and therefore ignore
the left-hand side of the $z$-component of momentum equation. Under these
assumptions, the surface
density $\Sigma$, the vertically integrated pressure $W$, and the half
thickness of the disk $H$ are given by

\begin{equation}
 \label{eq:def_sigma}
\Sigma =\int_{-\infty}^{\infty} \rho dz = \sqrt{2\pi}\rho_{0}(\varpi) H ~,
\end{equation}

\begin{equation}
 \label{eq:def_w}
W=\int_{-\infty}^{\infty} p_{\rm tot} dz = \frac{\Re T}{\mu}(1+\beta^{-1})\Sigma ~,
\end{equation}

\begin{equation}
 \label{eq:def_h}
H=\frac{1}{\Omega_{\rm K}}\left(\frac{W}{\Sigma}\right)^{1/2} =
\frac{1}{\Omega_{\rm K}}\left[ \frac{\Re T}{\mu} \left( 1 + \beta^{- 1}\right)\right]^{1/2} ~,
\end{equation}
where $\rho_{0}(\varpi)$ is the equatorial density and
$\Omega_{\rm K}=(GM/\varpi)^{1/2}/(\varpi-r_{\rm s})$
is the Keplerian angular speed. 

We now integrate other basic equations in the vertical
direction. Since the density vanishes at the disk surface, the vertically
integrated equation of continuity is expressed as

\begin{equation}
 \label{eq:int_continuity}
  \dot M = -2\pi\varpi \Sigma v_{\varpi} ~, 
\end{equation}
where $\dot M$ is the accretion rate. 

The $\varpi$-component of the
vertically integrated momentum equation can be obtained by using
$\langle {B_{\varphi}}^{2} \rangle = 8 \pi p_{\rm gas} {\beta}^{-1} = 8
\pi \beta^{-1} p_{\rm tot}/ \left( 1 + {\beta}^{-1}\right)$ ,

\begin{eqnarray}
 \label{eq:int_mom_pi}
  v_{\varpi} \frac{dv_{\varpi}}{d\varpi}+\frac{1}{\Sigma}
  \frac{dW}{d\varpi} & = & \nonumber \\
 \frac{\ell^2-\ell_{\rm K}^2}{\varpi^3}
 & - & \frac{W}{\Sigma} \frac{d{\rm ln}\Omega_{\rm K}}{d\varpi}
  - \frac{2}{\varpi} 
  \frac{\beta^{-1}}{1 + \beta^{-1}} \frac{W}{\Sigma} ~,
\end{eqnarray}
where $\ell_{\rm K}=\varpi^2 \Omega_{\rm K}$ is the Keplerian angular
momentum and the second term on the right-hand side is a correction
resulting from the fact that the radial component of the gravitational
force changes with height ( see \cite{mats84}).

Assuming that the last term of the $\varphi$-component of momentum equation
vanishes at the disk surface, the $\varphi$-component of vertically
integrated momentum equation is

\begin{equation}
 \dot M(\ell-\ell_{\rm in})=  2\pi \alpha_{\rm B} \varpi^2 W ~,
  \label{eq:int_mom_phi}
\end{equation}
where $\ell_{\rm in}$ is the specific angular momentum swallowed by the black
hole.

The vertically integrated energy equation is

\begin{equation}
Q_{\rm adv}=Q^{+} - Q^{-} ~,\label{eq:int_energy}
\end{equation}
where

\begin{equation}
Q_{\rm adv}=\frac{\dot M}{2\pi \varpi^2} \frac{\Re T}{\mu} \xi
\end{equation}
is the advective cooling rate, where

\begin{equation}
\xi = -\frac{3}{2}\frac{d{\rm ln}T}{d{\rm ln}\varpi}
+\frac{d{\rm ln}\Sigma}{d{\rm ln}\varpi}-\frac{d{\rm ln}H}{d{\rm
ln}\varpi} ~,
\end{equation}
and

\begin{equation}
 \label{eq:int_q+}
Q^{+} = \int_{-\infty}^{\infty} q^{+} dz =
  - \alpha_{\rm B} W \varpi \frac{d\Omega}{d\varpi} ~,
\end{equation}

\begin{equation}
 \label{eq:int_q-}
Q^{-} = \int_{-\infty}^{\infty} q^{-} dz = 6.2 \times
 10^{20} \rho_{0}^2 T^{1/2} \sqrt{\pi} H 
\end{equation}
are the heating and cooling rates integrated in the vertical direction.

\subsection{Prescription of the Advection Rate of the Toroidal Magnetic Flux}
The basic equations of the steady, magnetized accretion disks can be
closed by specifying the
radial distribution of magnetic field. By azimuthally averaging equation
(\ref{eq:vec_ind}), we obtain

\begin{eqnarray}
 \label{eq:ind}
 \frac{\partial \langle B_{\varphi} \rangle
 \mathbf{\hat{e}_{\varphi}}}{\partial t} & = & \nabla \times \left(
\mbox{\boldmath{$\bar{v}$}} \times \langle B_{\varphi} \rangle
\mathbf{\hat{e}_{\varphi}} \right) \nonumber \\
& + & \nabla \times \langle
 \delta \mbox{\boldmath{$v$}} \times \delta \mbox{\boldmath{$B$}}
 \rangle + \nabla \times
 \left( \eta \nabla \times \langle B_{\varphi} \rangle
  \mathbf{\hat{e}_{\varphi}}\right) ~,
\end{eqnarray}
where the second term on the right-hand side is the dynamo term and the
last term is the magnetic diffusion term. If we neglect these terms and
assume that the disk is in a steady state, we have

\begin{equation}
 \label{eq:steady_ind}
 0 = - \frac{\partial}{\partial z} \left(v_{z} \langle B_{\varphi}
				    \rangle  \right ) -
  \frac{\partial}{\partial \varpi} \left( v_{\varpi} \langle B_{\varphi}
				   \rangle \right) ~.
\end{equation}

Assuming that the azimuthally averaged toroidal magnetic fields vanish at
the disk surface, we can integrate equation (\ref{eq:steady_ind}) in the vertical direction and obtain,

\begin{equation}
 \label{eq:int_ind}
 \dot \Phi \equiv \int_{-\infty}^{\infty} v_{\varpi} \langle B_{\varphi}
 \rangle dz = constant ~.
\end{equation}

The radial advection rate of the toroidal magnetic flux $\dot \Phi$
(hereafter we call it flux advection rate) can be evaluated by completing
the vertical integration as

\begin{eqnarray}
 \label{eq:dot_phi}
\dot \Phi = - v_{\varpi} B_{0}(\varpi) \sqrt{4\pi} H ~,
\end{eqnarray}
where 
\begin{eqnarray}
 \label{eq:b0}
 B_{0}(\varpi) & = & \langle B_{\varphi} \rangle ( \varpi ; z = 0 )
 \nonumber \\
& = &
  2^{5/4} \pi^{1/4} (\Re T/\mu)^{1/2} \Sigma^{1/2} H^{-1/2}
  \beta^{-1/2}
\end{eqnarray}
is the azimuthally averaged equatorial
toroidal magnetic field. In accretion disks, $\dot \Phi$ is not always
conserved because $\dot \Phi$ can change with radius by the presence of
the dynamo term and the magnetic diffusion term in equation
(\ref{eq:ind}). 
According to \citet{mach06}, $\dot \Phi \propto {\varpi}^{-1}$ in the quasi
steady state. Based on this simulation result, we parametrize the
dependence of $\dot \Phi$ on $\varpi$ by introducing a parameter $\zeta$ as 
 
\begin{eqnarray}
\dot \Phi (\varpi ; \zeta , \dot M) \equiv \dot \Phi_{\rm out} (\dot M)
 \left(\frac{\varpi}{\varpi_{\rm out}}\right)^{-\zeta} ~,
 \label{eq:phiadv}
\end{eqnarray}
where ${\dot \Phi}_{\rm out}$ is the flux advection rate at the outer
boundary $\varpi=\varpi_{\rm out}$.
When $\zeta = 0$, the
magnetic flux is conserved. When $\zeta > 0$, the magnetic flux increases
with approaching to the black hole.

\section{Local Model}
\subsection{Local Approximation of Energy Equation}
Before obtaining global transonic solutions, we solve the energy equation
$Q^{+} = Q_{\rm adv} + Q^{-}$ locally 
at some specified radius to obtain thermal equilibrium curves of
optically thin black hole accretion flows.

We approximated the energy equation in order to be solved locally. We evaluated $\xi$ by using results of global three-dimensional
MHD simulations of black hole accretion flows. According to the results
by 
\citet{mach04}, $T \propto \varpi^{-1}$,  $\Sigma
\propto \varpi^{1/2}$ , and $H \propto \varpi$, thus $\xi =
1$. Furthermore, we assume $\Omega = \Omega_{\rm K}$. In our local
model, we adopt $\ell_{\rm in} = \ell_{\rm K} (3 r_{\rm s}) = 1.8371$,
$\varpi = 5 r_{\rm s}$ and $\alpha_{\rm B} = 0.05$ as the 
fiducial value, hence, the free parameters are now $\dot M$ and $\zeta$.

The flux advection rate $\dot
\Phi$ is determined from equation (\ref{eq:phiadv}) by specifying $\dot
\Phi_{\rm out}$. We assume that at $\varpi = \varpi_{\rm out} = 1000
r_{\rm s}$, $\ell_{\rm out}
= \ell_{\rm K}$, $T_{\rm out} = 
T_{\rm virial}$ and $\beta_{\rm out} = 10 $, where $T_{\rm virial} =
(\mu c^2 /3 \Re) (\varpi_{\rm out}/r_{\rm s})^{-1}$ is the virial
temperature at $\varpi = \varpi_{\rm out}$. Since we fixed $T$, $\beta$, 
and $\ell$ at $\varpi = \varpi_{\rm out}$, equations (\ref{eq:def_w})
,(\ref{eq:def_h}) and (\ref{eq:int_mom_phi}) indicate that $W_{\rm out}
\propto \Sigma_{\rm out}$, $H_{\rm out} \sim constant$ and $W_{\rm out}
\propto \dot M$, respectively. We find $\Sigma_{\rm out} \propto
\dot M$, thus, equation (\ref{eq:int_continuity}) indicates that
${v_{\varpi}}_{\rm out} \sim constant $. Therefore, equation (\ref{eq:b0})
gives $B_{0} (\varpi_{\rm out}) \propto {\Sigma_{\rm out}}^{1/2}
\propto {\dot M}^{1/2}$ and equation (\ref{eq:dot_phi}) gives ${\dot
\Phi}_{\rm out} \propto {\dot M}^{1/2}$.

\subsection{Numerical Results}
\begin{figure}[h]
 \begin{center}
  \includegraphics[width=80mm]{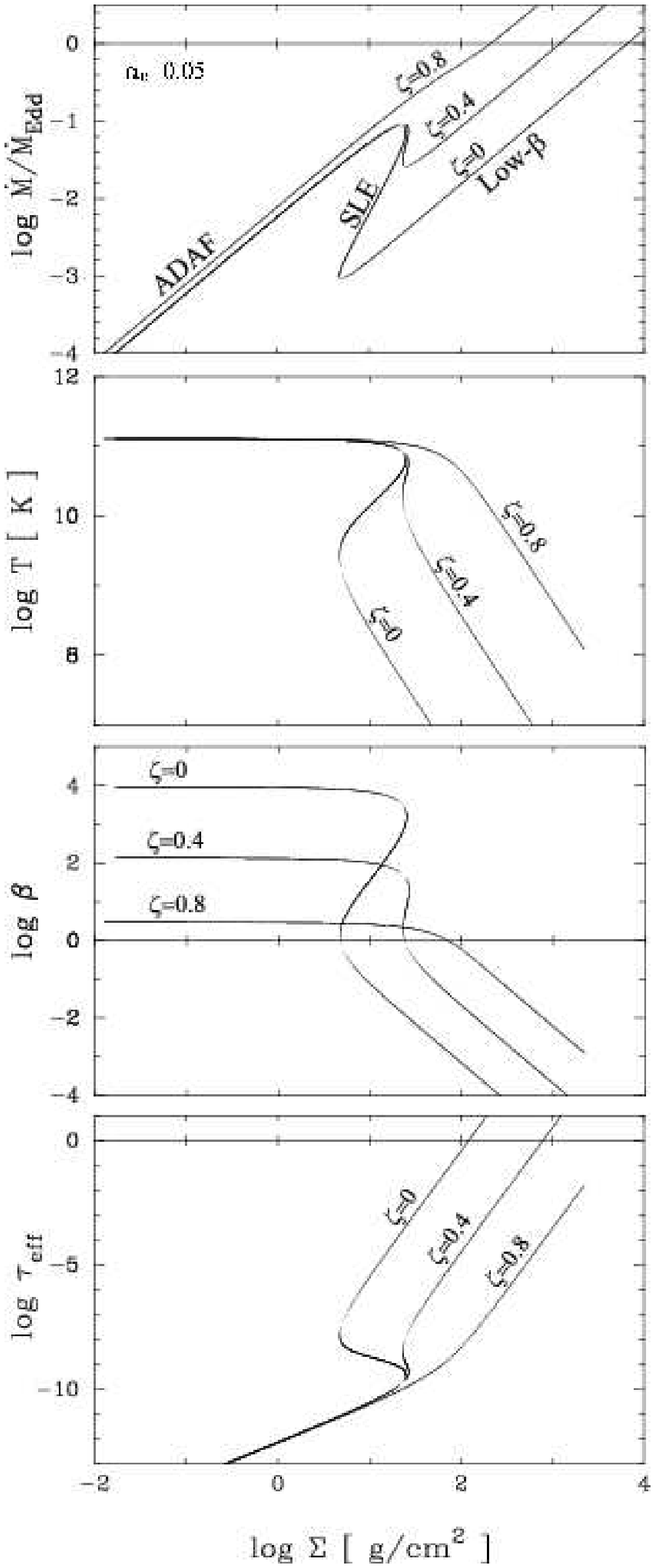}
  \caption{Local thermal equilibrium curves of accretion disks at
  $\varpi = 5 r_{\rm s}$ on the 
  $\Sigma - \dot M$, $\Sigma 
  - T$, $\Sigma - \beta$, $\Sigma - \tau_{\rm eff}$ plane for optically
  thin accretion disks with $M = 10 \MO$, $\alpha_{\rm B} = 0.05$ and $\zeta =
  0$, $0.4$, $0.8$. SLE indicates the cooling dominated branch by
  \citet{shap76}. ${\dot M}_{\rm Edd} = 4 \pi GM/(\eta_{\rm e}
  \kappa_{\rm es} c)$ is the Eddington accretion rate where the energy
  conversion efficiency $\eta_{\rm
  e}$ is taken to be $\eta_{\rm e} = 0.1$.
}
  \label{loal5}
 \end{center}
\end{figure}

In figure \ref{loal5}, we show the thermal equilibrium curves at $\varpi =5 r_{\rm
s}$ for optically thin accretion disks. The equilibrium curves are
plotted on the $\Sigma - \dot {M}$,
$\Sigma - T$, $\Sigma - 
\beta$ and $\Sigma - \tau_{\rm eff}$ plane for $\zeta = 0$, $0.4$ and
$0.8$. In the top panel of figure \ref{loal5}, $\dot M$ is normalized by the
Eddington accretion rate ${\dot M}_{\rm Edd} 
= 4 \pi G M / (\eta_{\rm e} \kappa_{\rm es} c) = 1.64 \times 10^{19} (0.1/
\eta_{\rm e}) (M/ 10\MO) ~ {\rm g} ~ {\rm s}^{-1}$, where the energy
conversion efficiency $\eta_{\rm e}$ and the electron scattering opacity
$\kappa_{\rm es}$ are taken to be $\eta_{\rm e} = 0.1$ and
$\kappa_{\rm es} = 0.34~  {\rm cm}^2 ~\rm {\rm g}^{-1}$, respectively. The
equilibrium curves  
consist of three branches; (1) gas pressure supported, advectively
cooled branch, which is thermally stable, i.e. ADAF branch, (2) gas
pressure 
supported, radiatively cooled branch, which is thermally unstable (SLE
branch obtained by \cite{shap76}), and (3) magnetic
pressure supported, radiatively cooled branch, which is thermally
stable. We call the last branch ``low-$\beta$ branch''. We find that the
low-$\beta$ branch exists even when $\dot M \gtrsim 0.1 {\dot M}_{\rm
Edd}$, which corresponds to $L \gtrsim 0.1 L_{\rm Edd}$.

It may be worth noting that on the low-$\beta$ branch, $\dot M \propto
\Sigma$ and $T \propto
{\Sigma}^{-2}$. This dependence can be explained as follows. When the magnetic
pressure is dominant, $W \sim {B_{0}}^2 H \propto (v_{\varpi} B_{0}
H)^{2} / ({v_{\varpi}}^{2} H) \propto {\dot \Phi}^{2} /({v_{\varpi}}^{2} H)
\propto \dot M / ({v_{\varpi}}^{2}
H)$. We find $v_{\varpi} \propto H^{-1/2}$ because equation
(\ref{eq:int_mom_phi}) gives $W \propto \dot M$ . Since equation
(\ref{eq:def_h})
gives $H \propto (W /\Sigma)^{1/2} \propto (\dot M / \Sigma)^{1/2}$ and
equation (\ref{eq:int_continuity}) gives $v_{\varpi} \propto \dot M / \Sigma$,
we obtain $\dot M \propto \Sigma$ and $H \sim
constant$. When the heating balances with the radiative cooling ( $Q^{+}
= Q^{-}$), since $Q^{+} \propto W$ 
and $Q^{-} \propto {\Sigma}^{2} T^{1/2} / H$, we find $W \propto \dot
M \propto {\Sigma}^{2} T^{1/2} / H$. Using $\dot M \propto \Sigma$ and $H \sim
constant$, we obtain $T \propto {\Sigma}^{-2}$.

The bottom panel of figure \ref{loal5} shows the effective optical depth
$\tau_{\rm eff} = \sqrt{\kappa_{\rm es} \kappa_{\rm ff}} ~ \Sigma / 2$
where $\kappa_{\rm ff} = 6.4 \times 10^{22} \rho_{0} (\varpi) T^{-3.5}
{\rm cm}^2 {\rm g}^{-1}$ is the opacity of free-free absorption. Note
that, when
$\zeta < 0.3$, the effective optical depth exceeds unity when 
$\dot M > 0.1 {\dot M}_{\rm Edd}$, thus, optically thin approximation is no
longer valid. We will extend the low-$\beta$ solution to optically thick
regime in subsequent papers.

One may wonder why $T$ and $\beta$ increase with $\zeta$ when $\Sigma$
is fixed in the low-$\beta$ branch in figure \ref{loal5}. When $\Sigma$
is fixed, since $W \sim  (B_{0} / 8\pi) H \propto {B_{0}}^{2} {W}^{1/2}$
on the low-$\beta$ branch, $Q^{+} \propto W \propto {B_{0}}^{4}$ increases
with $\zeta$. Thus, the equilibrium temperature increases according to
$W \propto Q^{+} = Q^{-} \propto T^{1/2} / H \propto T^{1/2} / W^{1/2}$
as $T \propto W^3 \propto {Q^{+}}^{3}$. The plasma $\beta$ on the
low-$\beta$ branch is given by $\beta \sim \Sigma T/W \propto T^{2/3}$
when $\Sigma$ is fixed. Thus, $\beta$ increases with $\zeta$.

Figure \ref{loal1} shows the thermal equilibrium curves for $\alpha_{\rm
B} = 0.01$. When $\alpha_{\rm B}$ is smaller, the transition from the ADAF
branch to the
low-$\beta$ branch takes place at smaller $\Sigma$ because heating
rate is smaller. The bottom panel of figure \ref{loal1} indicates that
as the accretion rate increases, the disk becomes optically thick before
the mass accretion rate attains $\dot M \sim 0.1 \dot M_{\rm Edd}$ when
$\zeta < 0.5$. 

In figure \ref{loal531r}, we show the radial dependence of the ${\dot
M}_{\rm crit}$ above which the ADAF branch disappears for $\zeta = 0.4$
and $\alpha_{\rm B} = 0.05$ (solid), $0.03$ (dashed), and $0.01$ (dotted). The
critical accretion rate ${\dot M}_{\rm crit}$ increases inward. This
indicates that as $\dot M$ increases the transition from ADAF branch to
low-$\beta$ branch takes place first at large radius and propagates
inward.

\section{Global Model}

We numerically solved equations (\ref{eq:int_continuity})-(\ref{eq:int_energy}) inward starting from the outer
boundary at $\varpi = \varpi_{\rm out}$ by specifying $\ell_{\rm out}$,
$T_{\rm 
out}$, $\beta_{\rm out}$, and four parameters $\alpha_{\rm B}$, $\dot M$,
$\zeta$, and $\ell_{\rm in}$. By adjusting the parameter $\ell_{\rm in}$, we
obtained a global transonic solution which smoothly passes the sonic
point (e.g., \cite{naka97}). The 
free parameters are now $\alpha_{\rm B}$, $\dot M$, and 
$\zeta$. We adopt $\alpha_{\rm B} = 0.05$.

We locate the outer boundary at $\varpi_{\rm out} =
1000 r_{\rm s}$, and imposed the boundary conditions $\ell_{\rm out} =
0.45 \ell_{\rm K}$, $T_{\rm out} = T_{\rm virial}$ and $\beta_{\rm out}
= 10$. We adopt backward-Euler method as an integration method (see \cite{mats84}).

In figure \ref{glal5r}, we show the radial structure of the disks when $\zeta = 1 $
and $\dot M /
{\dot M}_{\rm Edd} = 0.0002 $ (solid), $0.2446$ (dashed), $0.3181$
(dashed-dotted), and $0.4094$ (dotted). When $\dot M$ is small, since
$\Sigma$ is small in the whole disk, the radiative cooling is
inefficient. Thus the gas
does not cool down. The heating is balanced mainly with the advective cooling. Furthermore,
the flow is sub-Keplerian. When $\dot M$ is large, the disk can be
divided into 
four regions; the outer boundary region $\varpi \gtsim 400 r_{\rm s}$,
the 
outer transition region $400 r_{\rm s} \gtsim \varpi \gtsim 150 r_{\rm
s}$, the inner low-$\beta$ region $150 r_{\rm s} \gtsim \varpi \gtsim 50
r_{\rm 
s}$ and the inner ADAF region $\varpi \ltsim 50 r_{\rm s}$ when $\dot M/
{\dot M}_{\rm Edd} = 0.3181$ .
Near the outer boundary, the accretion flow is advection dominated;
$Q_{\rm adv} > Q^{-}$. However, in the outer transition region,
the radiative cooling overcomes the heating as $\Sigma$ increases. Below
this radius $T$ decreases inward, and $\beta$  
decreases, that is, the magnetic pressure increases inward due to the
magnetic flux conservation. At $\varpi \sim 100 r_{\rm s}$, $T$ ceases
to decrease inward because 
the heating balances with the radiative cooling. In the inner ADAF region, the advective cooling $Q_{\rm adv}$
increases as $T$ increases, 
and balances with the heating $Q^{+}$. The radiative
cooling decreases in the innermost region as $\Sigma$ decreases. 

To check the validity of the optically thin approximation, we show the
radial distribution of $\tau_{\rm
eff}$ in panel (f) of figure \ref{glal5r}. In these models, $\tau_{\rm eff}$ is
less than unity.

In the top panel of figure \ref{glal5}, we plotted the $\Sigma - \dot M$
relation at $\varpi = 5 r_{\rm s}$
obtained by global transonic solutions. We also computed the luminosity $L$ by 
integrating the radiative cooling all over the disk. In the bottom panel
of figure \ref{glal5}, we show the dependence of $L$, which is normalized by
the Eddington luminosity $L_{\rm Edd} = \eta_{\rm e} {\dot M}_{\rm Edd}
c^{2} = 1.47 \times 10^{39} ( M / 10 \MO) ~ {\rm erg} ~ {\rm s}^{-1} $, 
on the surface density at
$\varpi = 5 r_{\rm s}$. The optically thin low-$\beta$ solution extends
to $L \sim 0.1 L_{\rm Edd}$. Note that when $\Sigma \sim 10 ~{\rm g}~ {\rm
cm}^{-2}$
at $\varpi \sim 5 r_{\rm s}$ the disk luminosity is only $1 \%$ of the
Eddington luminosity even though the mass accretion rate attains $\dot M
\sim 0.1 {\dot M}_{\rm Edd}$. The disk is underluminous because the
innermost region of the disk is still advection dominated, that is,
radiatively inefficient. The dotted
curves in figure \ref{glal5} ($\zeta = 0.6$) show that when $10 ~ {\rm
g~ cm^{-2}}<
\Sigma < 100 ~ {\rm g~ cm^{-2}}$, $L$ increases although $\dot M$ is
nearly constant. It indicates that the disk luminosity increases as the
area of the radiative cooling dominated region increases.  The disk
luminosity can exceed $10 \%$ of the Eddington luminosity only when
$\dot M \gtrsim {\dot M}_{\rm Edd}$.

\section{Summary and Discussion}
We obtained steady transonic solutions of optically thin black hole
accretion flows by including magnetic fields. We theoretically
confirmed the simulation result by \citet{mach06} that
an optically thin hot disk undergoes a transition to an optically thin
magnetic pressure dominated disk when the accretion rate exceeds a
threshold. In addition to the ADAF branch and the SLE branch known for
optically thin disks, we found a low-$\beta$ branch in which the
magnetic 
heating balances with the radiative cooling.

In section 3, we obtained local solutions for single temperature disks
dominated by the toroidal magnetic fields. As a
consequence, we obtained three thermal equilibrium branches, ADAF, SLE
and low-$\beta$ branches.  

Based on the 
results of global three-dimensional MHD simulations by \citet{mach06},
we assumed that the Maxwell stress is proportional to the total
pressure, $\langle B_{\varpi} B_{\varphi} / 4 \pi \rangle  = -\alpha_{\rm B}
p_{\rm tot}$. Here $\alpha_{\rm B}$ is a parameter corresponding to the
$\alpha$-parameter in conventional theory of accretion disks. Note that
in magnetically dominated disks, since $p_{\rm tot} \sim p_{\rm mag}$,
the Maxwell stress is proportional to the magnetic pressure. In gas
pressure dominated disk, since $\beta \sim 10$ inside the disk except in
the plunging region of black hole accretion disks, $p_{\rm mag} \sim 0.1
p_{\rm tot}$. Thus our assumption is consistent with the results of
global three-dimensional MHD simulations by \authorcite{hawl01}
(\yearcite{hawl01}, \yearcite{hawl02}), in which they showed that
Maxwell stress is proportional to magnetic pressure (see also
\cite{pess06}). 

In this paper, we assumed that $\alpha_{\rm B}$ does not depend on $\beta$ and
is uniform for simplicity. 
Let us discuss the dependence of $\alpha_{\rm B}$ on the strength of
azimuthal magnetic fields. \citet{pess05} studied the stability of
accretion disks with superthermal azimuthal fields and showed that MRI
is suppressed when the Alfven speed inside the disk $v_{\rm A}$ exceeds
the geometrical mean of the sound speed $c_{\rm s}$ and the Keplerian
rotation speed $v_{\rm K}$ (i.e. $v_{\rm A} \ge \sqrt{c_{\rm s} v_{\rm
K}}$). Thus, the magnetic turbulence will be suppressed when this
criterion is satisfied. In the three-dimensional
MHD simulation reported by \citet{mach06}, fluctuating magnetic field
decreases as $\beta$ decreases to $\beta \sim 0.1$ (see Fig. 5b in
Machida et al. \yearcite{mach06}) because MRI is suppressed. Even in
such disks, however, the angular momentum is transported by the Maxwell
stress of mean magnetic fields, and $\alpha_{\rm B}$ remained
$\alpha_{\rm B} \sim 0.05 - 0.1$ (\cite{mach06}). We need to carry out simulations for
longer period to study the dependence of $\alpha_B$ on $\beta$. We
should note that $\alpha_{\rm B}$ increases in the innermost plunging
region of the disk (e.g., \cite{hawl02}; \cite{mach04}). Thus, this
dependence may affect on the global solutions passing through the sonic
point in the plunging region. However, its effect will be small for
local solutions outside the plunging region. We should also note that we
neglected the low-$\beta$ corona above accretion disks. It will be our
future work to take into account the effects of such corona and outflows
emerging from the disk. 

\begin{figure}[h]
 \begin{center}
  \includegraphics[width=80mm]{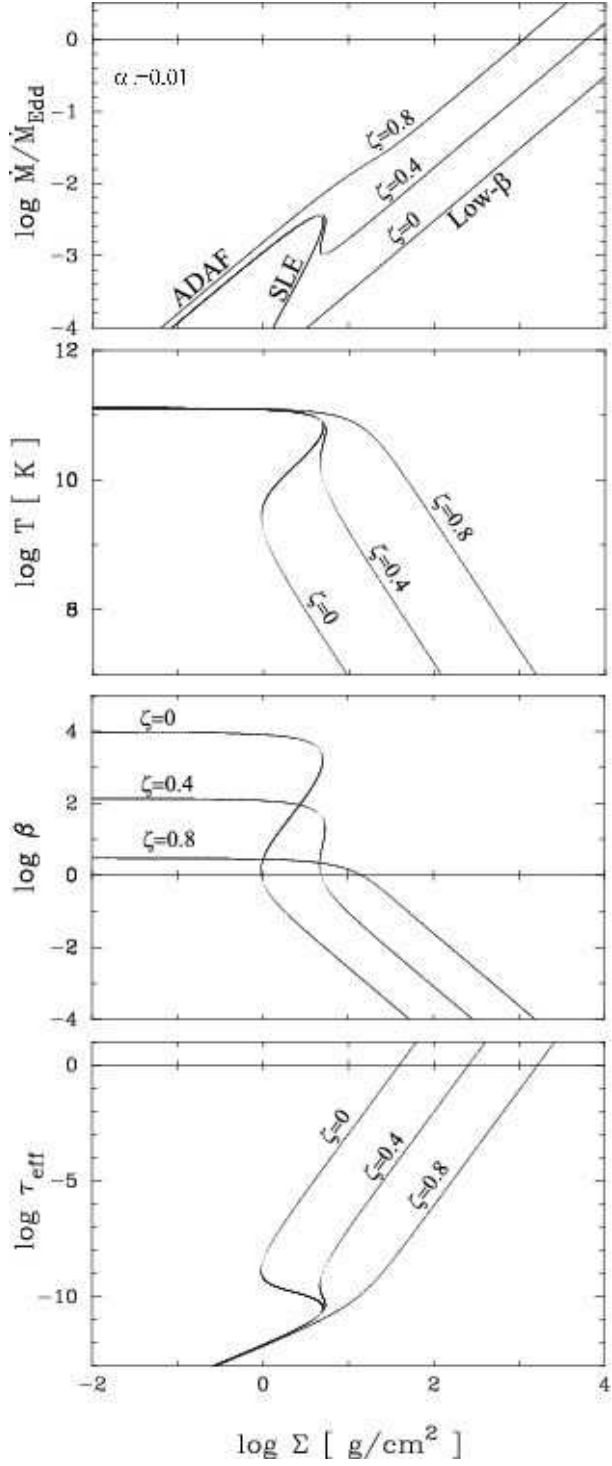}
  \caption{Same as figure \ref{loal5} but for $\alpha_{\rm B} = 0.01$.
}
  \label{loal1}
 \end{center}
\end{figure}

\begin{figure}[h]
 \begin{center}
  \includegraphics[width=80mm]{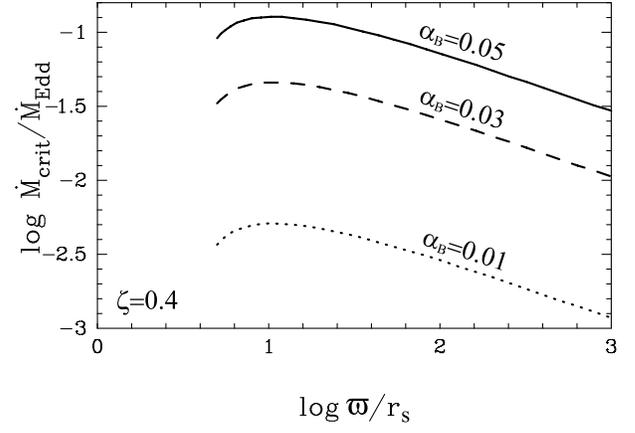}
  \caption{The radial dependence of ${\dot M}_{\rm crit}$ for $\zeta =
  0.4$ and $\alpha_{\rm B} = 0.05$ (solid), $0.03$ (dashed) and $0.01$
  (dotted). }
 \label{loal531r}
 \end{center}
\end{figure}

\begin{figure}[h]
 \begin{center}
  \includegraphics[width=160mm]{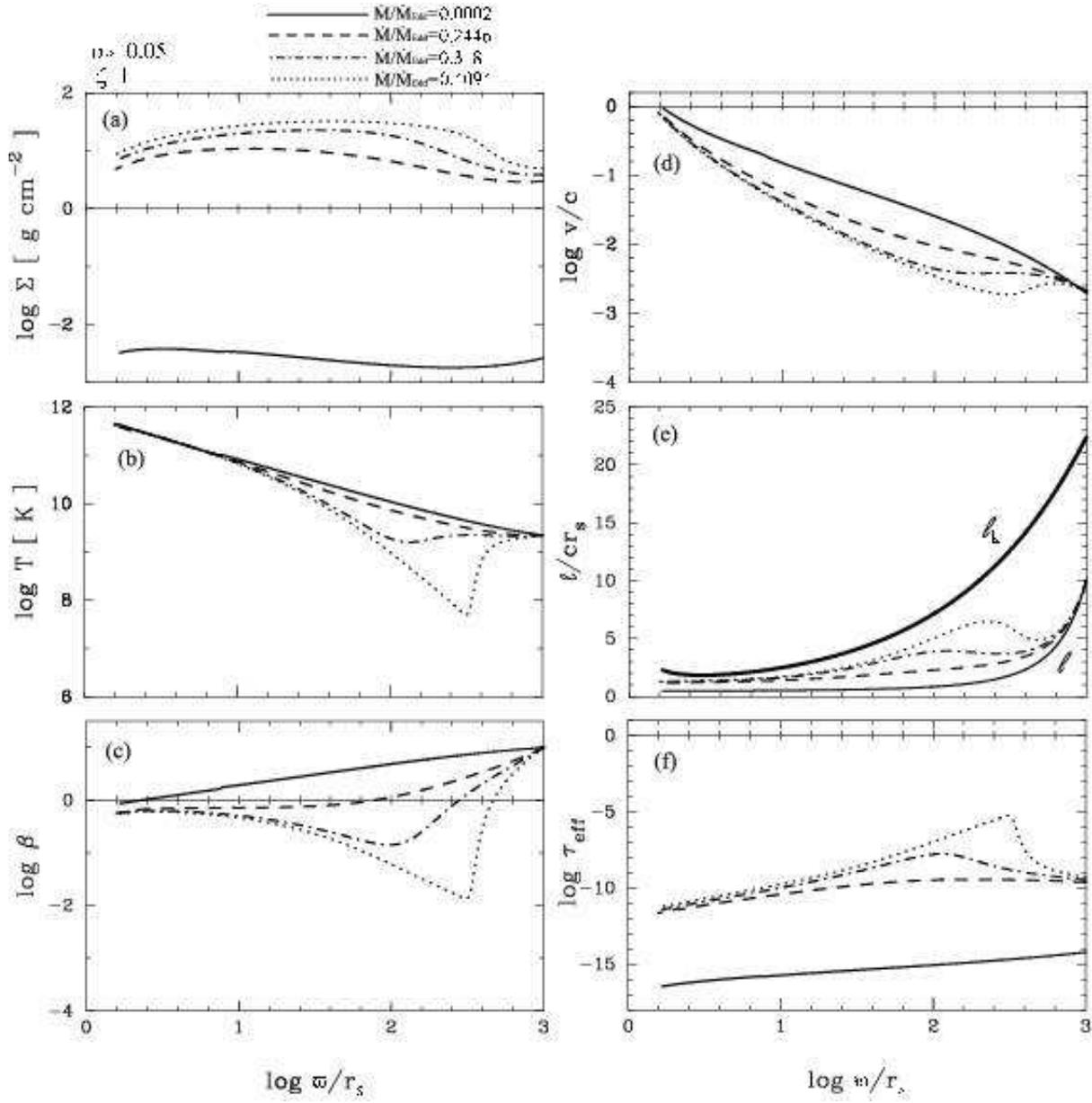}
  \caption{The radial structure of the disk when $\zeta = 1 $ and $\dot
  M /{\dot M}_{\rm Edd} = 0.0002 $ (solid), $0.2446$ (dashed),
  $0.3181$(dashed-dotted), and $0.4094$ (dotted) with $\alpha_{\rm B} =
  0.05$. Thick solid curve in panel (e) shows the Keplerian angular
  momentum distribution.
}
  \label{glal5r}
 \end{center}
\end{figure}

\begin{figure}[h]
 \begin{center}
  \includegraphics[width=80mm]{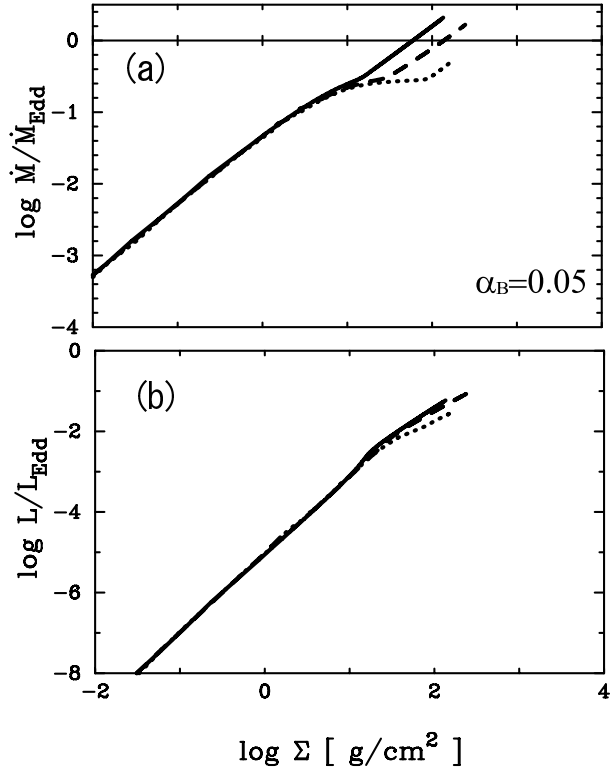}
  \caption{(a) $\Sigma - \dot M$ relation at $5
  r_{\rm s}$ obtained by global solutions with $\alpha_{\rm B} = 0.05$ when
  $\zeta = 1$ (solid), $0.8$ (dashed), 
  and $0.6$ (dotted). (b) The relation between $\Sigma$ and luminosity
  $L/L_{\rm Edd}$.
}
  \label{glal5}
 \end{center}
\end{figure}

The heating rate inside the disk
originates from the thermalization of magnetic energy via the magnetic
reconnection and is expressed as $q^{+} =
\langle B_{\varpi} B_{\varphi} / 4 \pi \rangle \varpi (d \Omega / d \varpi) = -
\alpha_{\rm B} p_{\rm tot} \varpi (d \Omega / d
\varpi)$. This expression is based on the three-dimensional MHD
simulations both in gas pressure dominant disks (\cite{hiro06}) and
in magnetic pressure dominant disks (Machida et al. \yearcite{mach06}). If MRI is suppressed
when $v_{\rm A} > \sqrt{c_{\rm s} v_{\rm K}}$ as \citet{pess05}
suggested, the dissipative heating rate $Q^{+}$ may decrease. Therefore,
the disk temperature will become smaller than that shown in figure
\ref{loal1}.

As we showed in section 3.2, $\dot M \propto
\Sigma$ and $T \propto {\Sigma}^{-2}$ on the low-$\beta$ branch. However,
this dependence is different from that given by \citet{mach06} in which
they derived 
$\dot M \propto {\Sigma}^{1/3}$ and $T \propto {\Sigma}^{-4}$. This difference
comes from the fact that in our model $ \dot \Phi \propto v_{\varpi} B H
\propto {\dot M}^{1/2}$ (that is, flux advection rate increases as mass
accretion rate increases). Meanwhile
\citet{mach06} assumed $B H \sim constant$.

The critical accretion rate for the ADAF solution $\dot M_{\rm crit}$
increases as 
decreasing the radius when $\varpi > 10 r_{\rm s}$, and has a maximum
around $\varpi \sim 10 r_{\rm s}$. This fact indicates that in the
region $\varpi > 10 r_{\rm s}$, the low-$\beta$ region propagates 
inward as $\dot M$ increases. When $\dot M$ exceeds the maximum of
${\dot M}_{\rm crit}$, whole disk will become low $\beta$.  

According to our local solutions, optically thin low-$\beta$ branches can exist above the
critical accretion rate ${\dot M}_{\rm crit}$ above which the ADAF
branch disappears (${\dot M}_{\rm crit} \sim 0.1 {\dot M}_{\rm Edd}$
when $\alpha_{\rm B} = 0.05$). The maximum luminosity of the disk
staying in the ADAF branch is only 10 \% of the Eddington
luminosity. To explain the observed X-ray hard state with luminosity 
as high as 20\% of the Eddington luminosity, the disk should be kept in
optically thin state. Such disks can exist when $\zeta \gtsim 0.5$ 
in our models, where $\zeta$ is the parameter which parameterizes the radial
variation of the advection rate of the toroidal magnetic flux.

We also obtained steady global transonic solutions including magnetic 
fields and obtained $\Sigma - \dot M$ relation. When $\zeta \sim 0$,
we could not obtain solutions above certain critical accretion rate. The
reason simply comes from the fact that we chose the ADAF-like outer
boundary condition. This means that there exists no solution connecting
from such outer boundary to inner region with weak magnetic field. When
$\zeta \sim 1$, we can obtain solutions for larger $\dot M$. These
solutions consist of four regions, the outer boundary region, the
outer transition region, the inner low-$\beta$ region and 
the inner ADAF region.

In this paper, we assumed single temperature plasma and included only
the bremsstrahlung cooling. In magnetized accretion disks, electrons
will also be cooled by synchrotron radiation and inverse Compton
effects. These processes will increase the luminosity of the disk and
decrease the critical accretion rate for the transition to the
low-$\beta$ disk.
In optically thin, hot accretion disks, since the gas temperature becomes
so high while electrons are cooled by radiation, the electron temperature
can become lower than the ion temperature. Such two-temperature
accretion disk models were studied by \citet{shap76} and
\citet{ichi77}. \citet{naka97} obtained transonic solutions of two
temperature accretion disks including synchrotron cooling and
inverse Compton effects. However, plasma $\beta$ was assumed to be
uniform in their paper. In subsequent papers, we would like to
obtain solutions of magnetized two temperature disks subjecting to the
cooling instability.

\bigskip
The authors thank Drs. M.A. Abramowicz, C. Akizuki, C. Done, J. Fukue, 
S. Hirose, Y. Kato, S. Mineshige, K. Ohsuga, R.A. Remillard, K. Watarai,
and S. Miyaji for their valuable comments and discussions. Part of this
work was carried out when R.M. and M.M. attended the KITP program on
"Physics of Astrophysical Outflows and Accretion Disks" at UCSB. This
work was supported in part by the Grants-in-Aid for Scientific Research
of the Ministry of Education, Culture, Sports, Science and Technology
(RM: 17030003), Japan Society for the Promotion of Science (JSPS)
Research Fellowships for Young Scientists (MM: 17-1907, 18-1907), and by
the National Science Foundation under Grant No. PHY99-07949.

\clearpage

\clearpage

\end{document}